\documentclass[twocolumn,aps,pra,showpacs,superscriptaddress]{revtex4}
\usepackage[usenames,dvipsnames]{color}

\usepackage[normalem]{ulem}
\usepackage{graphicx}
\usepackage{amsmath,amssymb}
\usepackage{color}
\usepackage{graphics}
\usepackage{epsfig}
\usepackage{bbm}
\usepackage{pstricks}

\newcommand{\be}{\begin{equation}}
\newcommand{\ee}{\end{equation}}
\newcommand{\eea}{\end{eqnarray}}
\newcommand{\bea}{\begin{eqnarray}}

\newcommand{\eins}{\openone}

\newcommand{\ketbra}[1]{\ensuremath{| #1 \rangle \!\langle #1 |}}

\newcommand{\ket}[1]{\ensuremath{|#1\rangle}}

\newcommand{\bra}[1]{\ensuremath{\langle#1|}}

\newcommand{\kommentar}[1]{}
\newcommand{\trace}{{\rm tr}}

\newcommand{\forget}[1]{}

\renewcommand{\a}[1]{\ensuremath{a_{#1}}}
\newcommand{\ad}[1]{\ensuremath{a_{#1}^\dagger}}

\renewcommand{\c}[1]{\ensuremath{c_{#1}}}
\newcommand{\bd}[1]{\ensuremath{b_{#1}^\dagger}}
\renewcommand{\b}[1]{\ensuremath{b_{#1}}}
\newcommand{\cd}[1]{\ensuremath{c_{#1}^\dagger}}
\newcommand{\ev}[1]{\langle #1 \rangle_{\vert 0 \rangle}}

\begin{document}


\title{Scaling of genuine multiparticle entanglement close to a quantum phase transition}

\author{Martin Hofmann}
\affiliation{Naturwissenschaftlich-Technische Fakult\"at,
Universit\"at Siegen,
Walter-Flex-Str.~3,
57068 Siegen, Germany}
\author{Andreas Osterloh}
\affiliation{Fakult\"at f\"ur Physik, Campus Duisburg,
Universit\"at Duisburg-Essen,
Lotharstr.~1, 47048 Duisburg, Germany}
\author{Otfried G{\"u}hne}
\affiliation{Naturwissenschaftlich-Technische Fakult\"at,
Universit\"at Siegen,
Walter-Flex-Str.~3,
57068 Siegen, Germany}

\date{\today}


\begin{abstract}
We investigate the scaling and spatial distribution of genuine multiparticle 
entanglement in three- and four-spin reduced states of the one-dimensional 
XY-model at the quantum phase transition. We observe a logarithmic divergence, show that genuine three- and four-particle entanglement obeys finite-size 
scaling of the XY-model and demonstrate that the genuine three-particle entanglement has a finite spatial range.
\end{abstract}



\maketitle


\section{Introduction}
In recent years it has turned out that quantum correlations play an 
important role in several branches of physics. One of these branches 
is the field of quantum optics and, more specifically, quantum 
information processing, where quantum correlations are often a 
useful resource for tasks like quantum-enhanced precision measurements 
or quantum key distribution. Another branch is the field of condensed 
matter physics: here, the study of entanglement in ground and thermal 
states of spin models has shed some light on the nature of quantum phase 
transitions \cite{Osterloh02, Osborne02, VidalLatorre03, Wu04, Roscilde04,Nataf12}, 
and in the same time has provided new insights into the simulation of 
spin models \cite{simulations}.

Most of the existing studies of entanglement in spin models, 
however, have restricted their attention to bipartite entanglement 
(for exemptions see \cite{Roscilde04, Patane07, multipartitespin, 
giampaolo, stasinska}): Either, the entanglement in the reduced 
state of two particles of a multiparticle system was studied 
\cite{Osborne02, Osterloh02}, or the entanglement between one 
block of particles and the remainder of the system \cite{VidalLatorre03}.  
In the latter case, the entanglement can often be related to the surface 
of the block, and these so-called area laws are an active research topic 
in mathematical physics \cite{Srednicki93, VidalLatorre03}, 
for a review see \cite{ECPlenio10}. The restriction to two-party entanglement 
is due to the fact that the theory of entanglement for multiparticle systems 
is still not fully developed, and many open problems exist. Therefore, in
the existing works on multiparticle entanglement  in spin models
\cite{multipartitespin, giampaolo, stasinska} only 
bounds on the entanglement are given. In a complementary approach Batle et al. \cite{batle} investigated nonlocality in spin chains using Mermin inequalities. The presence of nonlocality however, does not imply the presence of genuine multiparticle entanglement.

In this paper we study the scaling and spatial distribution of genuine multiparticle entanglement
at a quantum phase transition in one-dimensional spin models.  Our results
are enabled by recent progress in the theory of multiparticle entanglement 
\cite{Jungnitsch11,Kampermann12} in combination with an explicit determination of reduced
$k$-particle states in the transverse XY model \cite{Lieb61}. We consider 
a computable measure for genuine 
multiparticle entanglement for the reduced three- and four-particle states
and demonstrate that its derivative diverges at the critical point. For both 
cases we show that the entanglement obeys finite-size scaling, which can be 
used to compute the critical exponent for the infinite system from finite-size 
data.

\section{The model}
We consider the one-dimensional XY model with transverse 
magnetic field on $L$ particles and periodic boundary conditions 
\cite{Lieb61}. 
The Hamiltonian of this model is given by
\begin{equation}
H=-
\sum_{i=1}^L \frac{\lambda}{4}[(1+\gamma)\sigma_x^{(i)} \sigma_x^{(i+1)} + 
(1-\gamma) \sigma_y^{(i)} \sigma_y^{(i+1)}] + \frac{1}{2}\sigma_z^{(i)},
\end{equation}
where the coupling constant $\lambda \ge 0$ tunes the strength of the 
nearest neighbor coupling with respect to the external magnetic field. The 
parameter $\gamma$ sets the anisotropy of the system and connects the 
Ising model ($\gamma=1$) with the isotropic XY model ($\gamma=0$). 
In the thermodynamic limit and for $0<\gamma\le 1$ the ground state 
of the model undergoes a quantum phase transition at the critical 
point $\lambda_c=1$. For $\lambda=0$ there is a unique ground state, 
where all spins are aligned in the direction of the magnetic field 
and there is no magnetization in the $XY$-plane. For 
$\lambda\rightarrow\infty$ the ground state is two-fold degenerate
{and hence is an equal mixture of these two states}. 
At the quantum phase transition the systems ground state 
changes {from being non-degenerate to degenerate, 
accompanied by an abrupt change of the magnetization in 
the $x$-direction which is zero for $\lambda<1$ and finite 
for $\lambda \ge 1$.} We use the $XY$-model as a paradigm for our 
approach since it allows to study a phase transition with 
analytical rigor \cite{Lieb61}. 
In the following, we focus on the transverse Ising model ($\gamma=1$), 
but our approach can be straightforwardly extended to the $XY$-model.

\section{Genuine multiparticle negativity}
Our tool to study 
genuine multiparticle entanglement in the reduced marginals 
of the ground state of the spin model is the genuine multiparticle 
negativity \cite{Jungnitsch11}. This mixed state entanglement 
monotone vanishes on all biseparable states and a nonzero 
value is a sufficient criterion for genuine multiparticle 
entanglement to be present. Let us sketch its main idea and 
give the related definitions for three parties, Alice (A), Bob 
(B) and Charlie (C), the generalization to more parties is 
straightforward.

First, recall that a three-particle state is fully separable 
and contains no entanglement, if it can be written as a 
statistical mixture of product states, 
that is 
$\varrho^{\mathrm{fulsep}} = 
\sum_k p_k \ketbra{\psi^k_A} \otimes \ketbra{\phi^k_{B}} \otimes \ketbra{\phi^k_{C}}$. 
If the state is not fully separable, then it contains some 
entanglement, but 
it might be still separable with respect to one of the splittings 
$A\vert BC$, $B\vert AC$ or $C\vert AB$. 
In this case it can be written as statistical mixture of states 
on the first system and joint states on the last two systems, 
for example 
$\varrho^{\mathrm{sep}}_{A\vert BC} = 
\sum_k p_k \ketbra{\psi^k_A}\otimes\ketbra{\phi^k_{BC}}$. 
In this case, the state is called biseparable and not all 
three particles 
are entangled. Therefore, a state is called genuine multiparticle 
entangled \footnote{Sometimes the notion of ``genuine 
multiparticle entanglement'' is used in a more strict way, by 
requiring in addition that the state has a non-vanishing
value of some globally $SL$-invariant quantity \cite{OS05}. 
These states are discerned by states in the null-cone of global 
$SL$-invariance.}
if it is not a mixture of biseparable states, that is, 
it can {\it not} be written as 
\be
\varrho^{\mathrm{bisep}} = p_{A}\varrho^{\mathrm{sep}}_{A\vert BC} 
+ p_{B}\varrho^{\mathrm{sep}}_{B\vert AC} 
+ p_{C}\varrho^{\mathrm{sep}}_{C\vert AC}.
\ee
Clearly, for genuine multiparticle entangled states all particles are 
entangled and therefore this is the most interesting form of 
multiparticle entanglement 
\cite{gtreview}.

In order to study genuine multiparticle entanglement, it was proposed 
in Ref.~\cite{Jungnitsch11} to consider so-called PPT (positive partial 
transpose) mixtures. These are of the form 
\be
\varrho^{\rm PPTmix} = p_{A}\varrho^{\mathrm{PPT}}_{A\vert BC} 
+ p_{B}\varrho^{\mathrm{PPT}}_{B\vert AC} 
+ p_{C}\varrho^{\mathrm{PPT}}_{C\vert AC},
\ee
where $\varrho^{\mathrm{PPT}}_{A\vert BC}$ is a state for which the 
partial transpose with respect to the partition $A\vert BC$ has no 
negative eigenvalues \footnote{Formally, the partial transposition 
of a bipartite state 
$\varrho = \sum_{ij,kl} \varrho_{ij,kl} \ket{i}\bra{j}\otimes \ket{k}\bra{l}$
with respect to the second system is defined as 
$\varrho^{T_B}= \sum_{ij,kl} \varrho_{ij,kl} \ket{i}\bra{j}\otimes \ket{l}\bra{k}$.
}. 
Since separable states are also PPT \cite{pptcriterion}, the set of 
PPT mixtures is a superset of the biseparable states and hence all 
states which are not PPT mixtures are genuine multiparticle entangled. 
The advantage of the approach of Ref.~\cite{Jungnitsch11} is that the 
question whether a state is a PPT mixture or not can be decided directly, 
moreover, although PPT mixtures are only an approximation, the resulting 
entanglement criteria are the strongest criteria known. In fact, for many 
families of states (e.g. permutationally invariant three-qubit states), 
they are even necessary and sufficient for genuine multiparticle entanglement 
\cite{pptmixer}. 

The genuine multiparticle negativity $N_\varrho$ is a measure to 
distinguish between states that are PPT mixtures and such that are not. 
It vanishes, if a decomposition into a PPT mixture exists and is nonzero 
elsewhere. Moreover, it was proven to be an entanglement monotone, since 
it cannot increase under local operations and classical communication 
\cite{Jungnitsch11}. At this point, we mention that it can 
be computed directly for a given density matrix via semidefinite 
programming, the precise formulation is 
given in the Appendix \ref{appendix1}.

\begin{figure}
\begin{center}
\includegraphics[width=0.98\columnwidth]{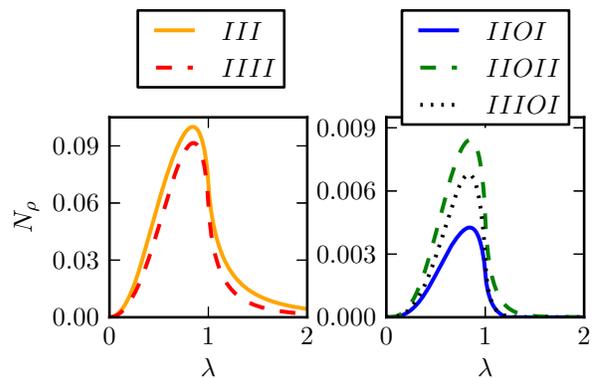}
\end{center}
\caption{The genuine negativity $N_\rho$ (for $L\rightarrow \infty)$
as a function of $\lambda$ for three- and four-particle reduced states 
in different spatial arrangements. On the left the values for three (III)  
or four (IIII) consecutive particles are shown. On the right hand 
side, spatial arrangements for three and four particles which contain a 
single vacancy (denoted by 0) are shown. Here, II0I denotes a three-particle 
reduced state in the qubits $\{i,i+1,i+3\}$ and II0II denotes a four-particle 
reduced state in the qubits $\{i,i+1,i+3,i+4\}$ etc.
See text for further details.}
	\label{fig:spacialgenneg}
\end{figure}

\section{Entanglement in reduced three-qubit states}
Let us start with 
the three-particle marginals of the ground state of the Ising model. 
Consider the particles $i$, $j$ and $k$ in the systems ground state. For 
the Ising model we can always set $i$ to be zero, since the $XY$-Hamiltonian 
is translationally invariant and has periodic boundary conditions. The 
spatial arrangement of the particles is then completely described by 
$(\alpha,\beta)=(j-i,k-j)$, where $(\alpha,\beta)$ and $(\beta,\alpha)$ 
lead to the same reduced states {due to mirror symmetry and hence one 
can choose $\alpha\le\beta$ without loss of generality}.

As a first task, we determined explicitly the reduced three-qubit states 
using the methods of Ref.~\cite{Lieb61}, detailed expressions are given in the 
Appendices \ref{appendix5} to \ref{appendix7}. Evaluating the genuine multiparticle negativity for different 
spatial constellations in the thermodynamic limit $L\rightarrow\infty$ 
we find that for $(1,1)$ and $(1,2)$ the reduced marginals of the ground 
state are genuine multiparticle entangled in the vicinity of $\lambda=1$ 
(see Fig.~\ref{fig:spacialgenneg}) and no entanglement is found if one 
separates the particles further.
As the criterion of PPT mixtures is 
in general only a sufficient criterion for entanglement, the question arises
whether for the separated configurations the reduced states are indeed biseparable.
Concerning this point, it it first worth mentioning that so far no example
of a genuinely entangled three-qubit state which cannot be detected by
the PPT mixture approach is known. In our case, we can show even explicitly, using 
the algorithm for proving separability from Ref.~\cite{Kampermann12} that 
the states are separable if the qubits are separated further (see Appendix 
\ref{appendix2}). Using these novel results we also conclude that the genuine 
three-particle entanglement in the ground state stays short ranged and falls 
off to zero. It answers a discussion recently raised in Ref.~\cite{giampaolo}:
In this reference, lower bounds on the entanglement were computed and no 
entanglement in the configuration $(1,2)$ was found, so it remained open, 
whether this or other configurations were separable. This makes us confident that the genuine multiparticle negativity is a well-suited tool for our analysis.

\section{Finite-size scaling}
The observed divergence indicates that the system undergoes a phase transition. 
We study it in more detail using the finite size scaling analysis \cite{fisher,barber}. 
This analysis is based on the idea that close to the phase transition at 
$\lambda_c$ the behavior of a diverging quantity $P^{(L)}(\lambda)$ 
for finite system sizes $L$ is governed by the system size $L$ and a 
rescaled variable $L/\xi$ only, with $\xi$ being the correlation 
length. In case of a logarithmic singularity of $P^{(L)}(\lambda)$, 
the finite-size scaling ansatz asserts the existence of a function $Q$, 
such that for finite $L$ and $\lambda$ close to the critical value \cite{barber}
\begin{equation}
P^{(L)}(\lambda)-P^{(L)}(\lambda_0) 
\sim 
Q(L^{\frac{1}{\nu}}\vert \lambda-\lambda_c\vert) 
- Q(L^{\frac{1}{\nu}}\vert \lambda_0-\lambda_c\vert).
\label{eqn:scalingansatz}
\end{equation}
Here $\nu$ is the critical exponent, which governs the divergence of the correlation 
length $\xi \sim \vert \lambda-\lambda_c\vert^{-\nu}$ close to the critical value 
as $L\rightarrow\infty$. For the ansatz to consistently recover 
\begin{equation}
P^{(\infty)}(\lambda) \sim C_\infty \ln \vert \lambda-\lambda_c \vert \text{ as }\lambda\rightarrow\lambda_c
\label{eqn:logdivquant}
\end{equation}
in the thermodynamic limit, one sets $Q(z)\sim C_\infty \ln z$ for $z\rightarrow \infty$. 
Provided that $Q(z)=const.$ as $z\rightarrow 0$
\begin{equation}
P^{(L)}(\lambda_c(L)) \sim -\frac{C_\infty}{\nu} \ln L + const.
\label{eqn:logdivmin}
\end{equation}
the minimum in $P^{(L)}$ at the pseudo-critical value diverges with the system size 
$L$, such that the equations (\ref{eqn:logdivquant}) and (\ref{eqn:logdivmin}) allow 
to determine the critical exponent $\nu$.

\begin{figure}
\begin{center}
\includegraphics[width=0.98\columnwidth]{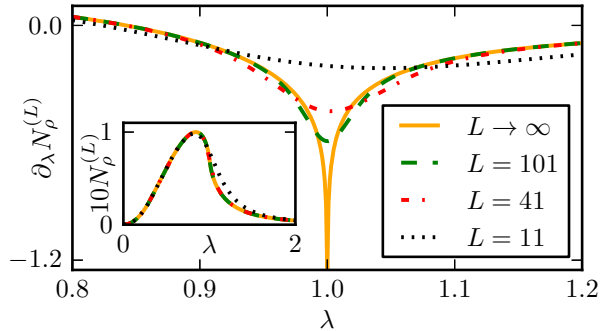}
\end{center}
\caption{In the thermodynamic limit the first derivative of the genuine 
negativity with respect to $\lambda$ diverges at the quantum phase 
transition. For finite chain length $L=11$ (dotted line), $L=41$ (dashed dotted line) 
and $L=101$ in (dashed line) there is a minimum in the vicinity of the 
critical point $\lambda_c=1$, which gets smaller for increasing $L$, 
diverging in the limit $L\rightarrow\infty$. The subplot shows the genuine 
negativity depending on $\lambda$ for different chain lengths in a 
larger region around the critical point.}
\label{fig:scalingderivative}
\end{figure}

For the two-particle entanglement, it was already shown that finite-size 
scaling holds \cite{Osterloh02}. In the multiparticle case, the quantity 
of interest is the derivative of the multiparticle negativity. 
So we vary the particle number $L$, keeping the arrangement $(1,1)$ 
fixed, and study the three-particle genuine negativity $N_\rho^{(L)}$ 
together with its first derivative $\partial_\lambda N_\rho^{(L)}$ (see Fig.~\ref{fig:scalingderivative}). 
One observes a logarithmic divergence of $\partial_\lambda N_\rho^{(\infty)}$ at 
$\lambda_c=1$. This is where the quantum phase transition occurs. There 
is distinct minimum in $\partial\lambda N_\rho^{(L)}$ for finite system 
sizes $L$ at $\lambda_c(L)$ which we take to be the pseudo-critical value. 
We find that it approaches the critical value like $\lambda_c(L)-\lambda_c \sim L^{-\kappa}$ with a shift exponent $\kappa=2.19$ \footnote{In many 
systems the shift exponent $\kappa$ equals the inverse of critical exponent 
$\nu$ of the diverging correlation length \cite{barber}. In our case, however, 
this turns out to be not the case.}.

\begin{figure}
\begin{center}
\includegraphics[width=0.98\columnwidth]{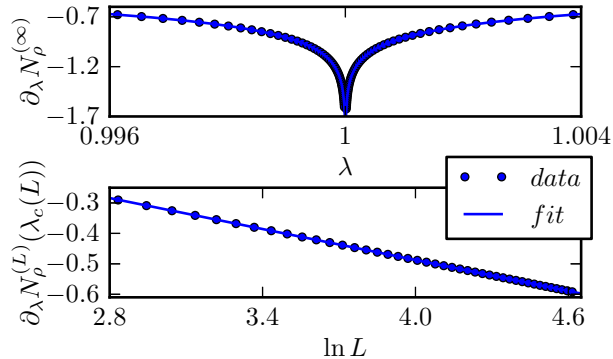}
\end{center}
\caption{Evaluation of the minima for different chain lengths and 
the divergence in the thermodynamic limit (circles) of the 
first derivative of the genuine negativity for three consecutive 
particles shows the behavior expected (lines) from the scaling 
ansatz for a logarithmic divergence. In the upper graph the 
divergence of the genuine negativity is plotted with respect 
to the coupling parameter $\lambda$ close to the critical 
point $\lambda_c=1$. The lower plot shows that minimum 
at the pseudo-critical value $\lambda_c(L)$ scales linear 
with the logarithm of the chain length $L$.}
\label{fig:scaling3particles}
\end{figure}

Fitting the 
expected behavior to our data as done in Fig.~\ref{fig:scaling3particles}, 
one recovers the critical exponent, which is known to be $\nu=1$, since
\begin{align}
	\partial_\lambda N^{(\infty)}_\varrho &= 0.170\ln (\vert\lambda-\lambda_c\vert) + 0.267\\
	\partial_\lambda N^{(L)}_\varrho(\lambda_c(L)) &= -0.170 \ln L + 0.191\; ,
\end{align}
for sufficiently big $L$. We performed the same analysis on the genuine negativity 
for the arrangements $(1,2)$ and observed similar qualitative results. Note that 
the numerical accuracy of the fitting procedure is higher than the displayed 
accuracy, a more detailed discussion is given in  Appendix \ref{appendix3}.

The finite-size scaling analysis shows that the behavior of the genuine 
multiparticle entanglement close to the critical point is governed by the 
quantum phase transition. A divergent behavior alone may be 
expected from the results of Ref.~\cite{Wu04}, as the reduced two-particle 
density matrix of the ground state itself is non-analytical \footnote{But 
note that the argument of Ref.~\cite{Wu04} in the strict sense concerns 
only entanglement in $k$-particle marginals of a ground state of an 
$k$-body Hamiltonian.}. Indeed many investigations of lower bounds on genuine multiparticle entanglement show this behaviour \cite{multipartitespin, giampaolo, stasinska}. On the flip side non of these investigations allowed to draw conclusions with respect to the critical attributes of the system.
 The finite-size scaling, however, shows that 
multiparticle entanglement faithfully represents important properties
of the spin system at the critical point, moreover, it may be used for
the extrapolation of critical exponents from finite-size numerical simulations.

\section{Entanglement in reduced four-qubit states}
As in the previous part, the density matrices of four 
particles $i<j<k<l$ depend on the spacing between the 
particles $\alpha=j-i$, $\beta=k-j$ and $\delta=l-k$ 
and the coupling parameter $\lambda$ only. Here we 
may choose $\alpha\le\delta$ due to symmetry.

We find that there are three spatial arrangements 
with non-zero genuine negativity. These are the tightest 
packed constellations $(1,1,1)$ and the constellations 
$(1,1,2)$ and $(1,2,1)$. Further separated constellations 
yield a zero genuine negativity. We observe that with 
increasing separation the four-particle genuine negativity 
decreases with increasing separation. This is similar to 
the three particle case. On the other hand, the genuine negativity 
for the four-particle arrangements with one intermediate 
particle $(1,1,2)$ and $(1,2,1)$ is much larger than the 
comparable three-particle case $(1,2)$ (see Fig. \ref{fig:spacialgenneg}). 
Taking into account that for the tightest conformations $(1,1)$ 
respectively $(1,1,1)$ the values are quite close to each 
other the four-particle genuine multiparticle entanglement 
seems to be more uniformly distributed throughout the system.

In addition,  we tested the four-concurrence, defined on pure states 
as $C_4(\psi):=\bra{\psi^*} \sigma_y\otimes \sigma_y\otimes \sigma_y\otimes \sigma_y\ket{\psi}$ and analytically extendable to mixed 
states \cite{Uhlmann00} on this model, and it qualitatively behaves 
roughly the same way, except for single distribution patterns. Since $C_4$ is non-zero only for products of two particle 
Bell states and $GHZ_4$ type of states, this observation suggests two
possible scenarios: either the state is a mixture of Bell products, not 
seen by $N_\rho$ plus states from the null-cone (for example $W_4$ states),
or it really contains $GHZ_4$ type entanglement. Further studies in this 
direction are needed in order to clarify this issue.

The finite-size scaling analysis for four consecutive particles 
$(1,1,1)$ yields results similar to the three-particle 
case but is more subject to numerical errors due to error propagation. 
For separations $(1,2,1)$ and $(1,1,2)$ the first derivative of 
the genuine negativity shows a qualitatively similar scaling behavior 
as in the case where all four particles are in succession. A full 
discussion is given in the Appendix \ref{appendix4}. 

\section{Discussion} 
Using the Ising model in a transverse magnetic field, we investigated 
the connection between genuine multiparticle entanglement and quantum 
phase transitions. We identified the configurations of three and 
four particles where entanglement is present and showed that the 
derivative of the genuine multiparticle negativity diverges logarithmically
at the critical points.  We further confirmed that this quantity
obeys a finite-size scaling behavior close to the quantum phase 
transition.

Besides its fundamental interest, there are several consequences and 
applications of our work. First, as our method allows the direct study 
of multiparticle entanglement, it can be used to complete existing 
indirect results on multiparticle entanglement. To give an example, in Ref.~\cite{Roscilde04} the existence of multiparticle entanglement in 
the one-dimensional XYZ model was concluded indirectly from monogamy 
relations, and possible connections to phase transitions were found. 
Our approach allows to verify these results in a direct manner. Similarly, 
in Ref.~\cite{Patane07} the two-partite negativities for different 
splittings of three particles in the XY model were studied, our 
methods can now decide whether these bipartite quantities are connected 
to genuine multiparticle entanglement. Further, in Ref.~\cite{stasinska} certain three-particle reduced could neither be detected as genuine multiparticle entangled nor shown to be biseparable. We expect that an application of our method would solve this issue.
 Second, our results demonstrate 
the usefulness of the genuine multiparticle negativity to study many-body 
systems. This makes it applicable to further systems, such as dynamical 
phase transitions \cite{Sen05}, quenching dynamics, or the study of
symmetry breaking \cite{stasinska}.

\section*{Acknowledgments}
We thank Steve Campbell, Tobias Moroder and Anna Sanpera for discussions. This work
 has been supported by the EU (Marie Curie CIG 293993/ENFOQI) and the
 BMBF (Chist-Era Project QUASAR). A.O. acknowledges the financial 
support by the SFB TR12.

\appendix

\begin{widetext}
\section{}

\subsection{Mathematical definition of the multiparticle negativity}
\label{appendix1}
Mathematically, the genuine multiparticle negativity is defined as 
optimization over the set of fully decomposable witnesses $W$, these 
witnesses have to fulfill certain constraints \cite{Jungnitsch11}. In total, 
the optimization problem reads for three particles:
\begin{align}
\nonumber	N_\varrho & = -\min_W\ \trace(W\varrho)\\
\nonumber	&\text{such that for all } m\in \left\{ A,B,C \right\}\\
		&W=P_m+Q_m^{T_m},\ 0\le P_m,Q_m\le \eins.
\end{align}
As all these conditions are linear or semidefinite constraints, this problem can be
solved with the method of semidefinite programming, and the optimality of the 
solution can be verified. A simple, ready-to-use implementation can be found 
on-line \cite{pptmixerprogram}.

\subsection{Separability of three and four qubits}
\label{appendix2}
In  our work we supplement the entanglement monotone with a separability algorithm \cite{Kampermann12}. The idea of the algorithm is to decompose $\varrho$ into two biseparable parts 
\begin{equation}
	\varrho=(1-p)\varrho_a+p\varrho_b,
	\label{eqn:dec}
\end{equation}
such that $\varrho_b$ is a statistical mixture of pure separable states and $\varrho_a$ is within the ball of separable states around the completely mixed state \cite{Gurvits02}. The algorithm is performed iteratively such that the purity of $\varrho_a$ is decreased with each successful step until $\varrho_a$ lies within the set of separable states proving the separability of $\varrho$.

Given an input state $\varrho$, we define $\varrho=\varrho_0$ and iterate the following procedure:
\begin{itemize}
	\item Find a pure biseparable state $\ket{\psi_k}$ which has large overlap with $\sqrt{\varrho_k}$. It will add to the part $\varrho_b$ in the decomposition (\ref{eqn:dec}). We take random states and choose the one with largest overlap.
	\item Choose $0 \le \varepsilon_k \le 1$, such that $\varrho_{k+1} = \frac{1}{1-\varepsilon_k}(\varrho_k-\varepsilon_k\ketbra{\psi_k})$ is positive semidefinite. This ensures, that $\varrho$ is indeed a convex combination of all $\ketbra{\psi_k}$ and $\varrho_{k+1}$. Note that the algorithm seems to be more reliable, if $\varepsilon_k$ is significantly smaller than the smallest eigenvalue of $\varrho$.
	\item Check whether $tr({\varrho_{k+1}}^2)<\frac 1 7$ and thus if $\varrho_{k+1}$ is separable for some bipartition \cite{Gurvits02}.
	\item If this is the case, than the algorithm finishes and $\varrho$ is separable, else continue until some maximal iteration number is reached. 
\end{itemize}
In practice the algorithm performs good on all full rank states, if however the state has eigenvalues close to zero it fails to detect separable states as such. This is due to the fact, that $\rho_0$ is initially close to the border of separable states and hence during iterations $\rho_{k}$ might leave the set of separable states. In such a case the algorithm can not bring back $\rho_k$ into the separable ball around the completely mixed state and the algorithm does not detect the state as separable.

Using local filtering one can significantly improve the algorithms ability to detect separable states with small eigenvalues. The idea is to apply invertible local matrices to the state $\varrho \mapsto \mathcal F_1 \otimes \mathcal F_2 \otimes \mathcal F_3 \varrho \mathcal F^\dagger_1 \otimes \mathcal F^\dagger_2 \otimes \mathcal F^\dagger_3$, such that the smallest eigenvalue of the normalized new state increases. This new states is then more likely to be detected by the algorithm if it is separable. Since both states are connected by invertible local operations the separability of one state implies the separability of the other. Using this modification we can on the one hand decrease the number of iterations the algorithm needs to show the separability of a separable state. On the other hand we are able to detect separable states, which are not detected by the unmodified algorithm.

Our investigations of the three-particle reduced states in the thermodynamic limit for the particle separation other than $(1,1)$ and $(1,2)$ result in a vanishing genuine multiparticle negativity. Applying the algorithm to states of the constellations $(1,3)$ and $(2,2)$ confirms that these reduced three-particle marginals are separable (see Fig. \ref{fig:separabilityalgorithm}).

\begin{figure}[t]
	\begin{center}
		\includegraphics[width=0.4\textwidth]{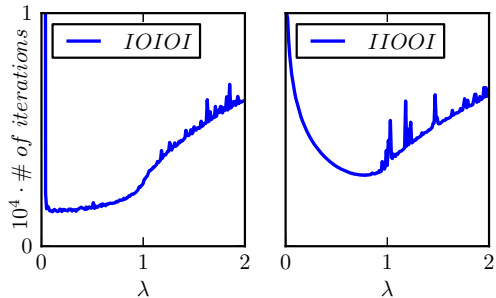}
	\end{center}
	\caption{The number of iteration steps of the separability algorithm needed to show separability is shown with respect to the coupling parameter $\lambda$ for the particle constellations $(1,3)$ (right) as well as for $(2,2)$ (left). For most cases the algorithm converges in less than $10^4$ steps proving the separability of the respective states. For $\lambda=0$ we know that the reduced state is pure and separable. For $\lambda \rightarrow 0$ the underlying state has small eigenvalues. Here the algorithm does not prove separability, even though these states are separable as well.}
	\label{fig:separabilityalgorithm}
\end{figure}

\subsection{Four successive particles scaling analysis}
\label{appendix3}
The scaling analysis in the case of four closely packed particles $(1,1,1)$ is along the line of the three-particle case. It yields similar results as in the three-particle case (see figure \ref{fig:scaling4particles})
\begin{align}
	\partial_\lambda N^{(\infty)}_\varrho &= 0.20\ln (\vert\lambda-\lambda_c\vert) + 0.36\\
\partial_\lambda L^{(L)}_\varrho (\lambda_c(L)) &= -0.20 \ln L + 0.27.
\end{align}
The negative quotient of the logarithmic prefactors again give way to the expected critical exponent of the diverging correlation length $\nu=1$. Due to numerical precision, however, the determination of the position and value of the minima was more subtle than in the tree particle case. Especially a faithful determination of the position of the minima was impossible for $L>33$ due to numerical inaccuracies, which we discuss in the next section in this appendix. The determination of the absolute value of the minima however suffered less from these problems and made it possible to confirm the expected scaling.

\begin{figure}[t]
	\begin{center}
		\includegraphics[width=0.4\textwidth]{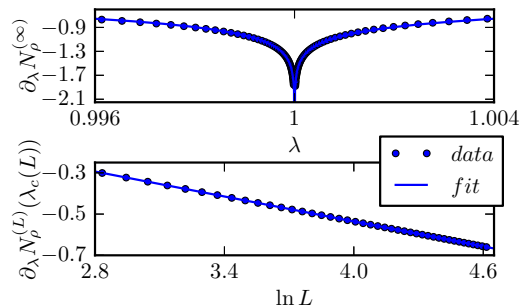}
	\end{center}
	\caption{Evaluation of the minima for different chain lengths and the divergence in the thermodynamic limit (circles) of the first derivative of the genuine negativity for four consecutive particles shows the scaling behavior expected (lines) from the scaling ansatz for a logarithmic divergence. On top the genuine negativity is plotted with respect to $\lambda$ diverging logarithmically at the critical point. The mid plot confirms the ansatz made with respect the $L$-dependence of position of the minima and the lower plot shows, the predicted behavior of the value of the minima with respect to different chain length.}
	\label{fig:scaling4particles}
\end{figure}

\subsection{Discussion of numerical precision}
\label{appendix4}
For the derivation of our results we make use of the genuine negativity \cite{Jungnitsch11}, which is implemented via semidefinite programming. It provides an upper bound of how far the numerical value obtained is from the global optimum. One can take this bound as a conservative error for the genuine negativity. In our calculations the absolute error of the genuine negativity is of the order of $10^{-14}$. This error is negligible, if one considers the genuine negativity itself. For the finite-size scaling analysis however this error has to be taken into account. Recall, that in the scaling analysis one has to extract several minima of the first derivative of the genuine negativity. The first derivative is approximated with the central finite difference method
\begin{equation}
	f' \approx \frac{-f(x+2h)+8f(x+h)-8f(x-h)+f(x-2h)}{12h}
\end{equation}
and the forward and backward finite differences close to the divergence
\begin{align}
\nonumber	f' & \approx \frac{-25f(x)+48f(x+h)-36f(x+2h)+16f(x+3h)-3f(x+4h)}{12h}\\
		f' & \approx \frac{+25f(x)-48f(x-h)+36f(x-2h)-16f(x-3h)+3f(x-4h)}{12h}.
\end{align}
We found $h=10^{-7}$ to give us optimal results. The absolute error of the genuine negativity, results in an absolute leading error in the first derivative. It is of the order of $10^{-7}$ and hence the overall precision in estimating the position and absolute value of the minimum in the first derivative of the genuine negativity is limited. This directly influences how well one can perform the finite-size scaling analysis. In case of three $(1,1)$ and four  $(1,1,1)$ consecutive particles one can find the position of these minima with sufficient high precision up to $L=33$ particles and their values up to $L=100$ particles, which is sufficient for the finite-size scaling analysis. For a larger number of particles however the errors of the positions and values of the minima caused growing errors in the coefficients of the fitting functions. This finally makes the extraction of the critical exponent $\nu$ impossible. Hence, we omitted a quantitative finite-size scaling analysis in these cases.

\subsection{Diagonalizing the XY model}
\label{appendix5}
We give a brief review of the analytical diagonalization of
\begin{equation}
	H=-
	\sum_{i=1}^L \frac{\lambda}{4}[(1+\gamma)\sigma_x^{(i)} \sigma_x^{(i+1)} + 
	(1-\gamma) \sigma_y^{(i)} \sigma_y^{(i+1)}] + \frac{1}{2}\sigma_z^{(i)}.
	\label{eqn:HXY}
\end{equation}
The strategy is as follows \cite{Lieb61}.
\begin{itemize}
\item First, we apply the so called Jordan-Wigner transformation, which maps the model onto a fermionic Fock space with creation and annihilation operators.
\item Second, the transformed system decouples into a direct sum of four dimensional Hilbert spaces by using discrete Fourier transformation.
\item Finally, these subspaces can be diagonalized, giving access to the ground state, the energy spectrum and expectation values of finite products of one site Pauli operators.
\end{itemize}
First, apply the Jordan-Wigner transformation, which transforms the spin operators $S^j_\alpha$, $j=1,\dots,L$, $\alpha=x,y,z$ into fermionic creators $\cd{j}$ and annihilators $\c{j}$. It is composed of two intermediate transformations
\begin{equation}
 \ad{j} = S_x^j+iS_y^j \text{ and } \a{j} = S_x^j - i S_y^j,
\end{equation}
where $\ad{j}$ and $\a{j}$ are hard-core bosonic creation and annihilation operators. The fermionic operators are then obtained by
\begin{equation}
 \c{j} = \exp\left\{ \pi i \sum_{k=1}^{j-1}\ad{k}\a{k} \right\}\a{j}, \ \ \ \cd{j}= \ad{j} \exp\left\{ -\pi i \sum_{k=1}^{j-1}\ad{k}\a{k} \right\}.
\end{equation}
The new operators obey the fermionic anti-commutation algebra
\begin{equation}
\left\{ \cd{i},\c{j} \right\} = \delta_{ij} \text{ and } \left\{ \c{i},\c{j} \right\}=\left\{ \cd{i},\cd{j} \right\}=0.
\end{equation}
Applying the Jordan-Wigner transformation to the Hamiltonian (\ref{eqn:HXY}) yields
\begin{equation}
	H = \frac{L}{2} + \frac{\lambda}{2} \left[ \left( \cd{L}\c{1}+\gamma \cd{L}\c{1}\right) + h.c. \right]\left( \exp\left\{ i\pi\sum_{j=1}^L\cd{j}\c{j} \right\}+1 \right) - \frac{\lambda}{2} \sum_{i=1}^L \left( \cd{i}\c{i+1}+\gamma\cd{i}\cd{i+1} \right) + h.c. + \cd{i}\c{i},
\end{equation}
where $h.c.$ denotes the hermitian conjugate of the parenthesized expression in front. For $L$ large the second term can be neglected. For odd $L$ the second term vanishes. Thus by restriction to an odd number of particles, one can proceed the diagonalization procedure with the Hamiltonian
\begin{equation}
	H = \frac L 2 - \sum_{i=1}^L \frac{\lambda}{2} \left( \cd{i}\c{i+1}+\gamma\cd{i}\cd{i+1} \right) + h.c. + \cd{i}\c{i}.
\label{eqn:XYHamiltonianfermionic}
\end{equation}
Next, one performs a Fourier transform to decouple the Hamiltonian into a direct product of Hilbert spaces preserving the fermionic anti-commutation relations. Let $\phi_p = 2\pi p/L$, then the operators $\c{i}$ are given in the new operators $\b{p}$ as
\begin{equation}
	\label{eqn:fouriertrafo} \c{j} = \frac{1}{\sqrt{L}} \sum_{p=-L/2}^{L/2} \exp(-ij\phi_p)\b{p},
\end{equation}
where the summation runs over all possible momenta. Applying the Fourier transform (\ref{eqn:fouriertrafo}) to the Hamiltonian (\ref{eqn:XYHamiltonianfermionic}) results in
\begin{equation}
	H = \frac L 2 -(\lambda+1)\bd{0}\b{0} - \sum_{p=1}^{L/2} (\lambda \cos \phi_p + 1)(\bd{p}\b{p}+\bd{-p}\b{-p}) - i\gamma \lambda\sin\phi_p(\bd{p}\bd{-p}+\b{p}\b{-p}).
	\label{eqn:quadraticformofsubspacehamiltonian}
\end{equation}
The Hilbert space decomposes into non-interacting subspaces. The space of  zero momentum is two dimensional and already diagonal, whereas the other subspaces are four dimensional and non diagonal.

In order to diagonalize the remaining subspaces in the Hamiltonian (\ref{eqn:quadraticformofsubspacehamiltonian}) let $\alpha_p=(\lambda\cos\phi_p+1)$, $\beta_p=\lambda\gamma\sin\phi_p$ and use then the canonically transformed operators
\begin{equation}
	\eta_k = \tilde{\alpha}_k\b{-k} - i\tilde{\beta}_k\bd{k},
\end{equation}
with
\begin{align}
\nonumber	\tilde{\alpha}_k &= \frac{\Lambda_k-\alpha_k}{\sqrt{2\left( \Lambda_k^2 -\Lambda_k\alpha_k \right)}},\\
\nonumber	\tilde{\beta}_k &= \frac{\beta_k}{\sqrt{2\left( \Lambda_k^2 -\Lambda_k\alpha_k \right)}} \text{ and}\\
		\Lambda_k &= \sqrt{\alpha_k^2+\beta_k^2}.
\end{align}
In the new creation and annihilation operators $\eta_k^\dagger$ and $\eta_k$ our Hamiltonian reads
\begin{equation}
	H = \sum_{k=-L/2}^{L/2} \Lambda_k \eta_k^\dagger \eta_k - \frac{1}{2}\sum_k\Lambda_k.
	\label{eqn:diagonalizedhamiltonian}
\end{equation}
Clearly, the ground state of the system is given by the Fock vacuum in the fermionic basis. Although the vacuum state is separable in the fermionic basis this does not have to hold for the computational basis, since the Jordan-Wigner transformation is a global unitary transformation. To draw conclusions about the entanglement properties of the ground state one has to study the ground state in the computational basis.

\subsection{Expectation values of Pauli operators}
\label{appendix6}
In order to calculate the three-particle \cite{Patane07} respectively four-particle reduced density matrices of the  ground state $\ket{0}$ of our system one has to trace out all particles but three, respectively four. Alternatively one can choose a local operator basis on each system, which is not to be traced out and calculate the one-point up to four-point correlators to recover the reduced density matrix. For the reduced states on the tree particles $i$, $j$ and $k$ and the four particles $i$, $j$, $k$ and $l$ this would yield
\begin{align}
\label{eqn:treeparticlesrdeducednosymm}	\varrho_{ijk} &= \frac 1 8 \sum_{m,n,o } \ev{\sigma_i^m\sigma_j^n\sigma_k^o}\sigma_i^m\sigma_j^n\sigma_k^o \text{ and}\\
\label{eqn:fourparticlesrdeducednosymm}	\varrho_{ijkl} &= \frac{1}{16} \sum_{m,n,o,p} \ev{\sigma_i^m\sigma_j^n\sigma_k^o\sigma_l^p}\sigma_i^m\sigma_j^n\sigma_k^o\sigma_l^p,
\end{align}
where each of the summations $m$, $n$, $o$ and $p$ run over $\left\{ x,y,z,0 \right\}$. In order to calculate the expectation values $\ev{\sigma_{i_1}^{\alpha_1}\dots\sigma_{i_k}^{\alpha_k}}$ one needs to express the Pauli-operators in terms of the fermionic operators $\eta_k^\dagger$ and $\eta_k$. It is
\begin{align}
\nonumber	\sigma_l^x &= (\c{l}+\cd{l})\prod_{i=1}^{l-1}(\cd{i}+\c{i})(\cd{i}-\c{i}),\\
\nonumber	\sigma_l^z &= -(\cd{l}+\c{l})(\cd{l}-\c{l}),\\
		\sigma_l^y &= -i (\cd{l}-\c{l})\prod_{i=1}^{l-1}(\cd{i}+\c{i})(\cd{i}-\c{i})
\end{align}
and thus the Pauli operators are products of
\begin{align}
\label{eqn:Al} A_l &= \c{l}+\cd{l} = \frac{1}{\sqrt{L}} \sum_q \left( \eta_{-q}^\dagger + \eta_q \right)(\alpha_q+i\beta_q)e^{i\phi_q l},\\
\label{eqn:Bl} B_l &= \c{l}-\cd{l} =  \frac{1}{\sqrt{L}} \sum_q \left( -\eta_{-q}^\dagger + \eta_q \right)(\alpha_q-i\beta_q)e^{i\phi_q l},
\end{align}
which are linear in the creation and annihilation operators $\eta_i^\dagger$ resp. $\eta_i$. The expectation values of arbitrary tensor products of
\begin{align}
\nonumber	\sigma_l^x &= A_l\prod_{i=1}^{l-1}A_iB_i,\\
\nonumber	\sigma_l^z &= -A_iB_i,\\
		\sigma_l^y &= -i B_l\prod_{i=1}^{l-1}A_iB_i,
\end{align}
are then monomials in the fermionic creation and annihilation operators. These can be evaluated using the Wick theorem, which states the equality
\begin{equation}
\label{eqn:wickthm}
\ev{\mathcal O_1\cdots \mathcal O_n} = \ev{\mathcal O_1\mathcal O_2}\ev{\mathcal O_3\cdots \mathcal O_n} - \ev{\mathcal O_1 \mathcal O_3}\ev{\mathcal O_2\mathcal O_4\cdots \mathcal O_n}+\ev{\mathcal O_1 \mathcal O_4}\cdots,
\end{equation}
where $\mathcal O_i$ can be any operator $A_j$ or $B_k$. Successive application of the theorem reduces the expectation values in (\ref{eqn:treeparticlesrdeducednosymm}) and its four-particle counterpart (\ref{eqn:fourparticlesrdeducednosymm}) to products of the two-point expectation values $\ev{ A_lA_k}$, $\ev{A_lB_k}$ and $\ev{B_lB_k}$. Using equations (\ref{eqn:Al}) and (\ref{eqn:Bl}) and setting $r=k-l$ one can calculate these to be
\begin{align}
\label{eqn:wick1}\ev{ A_lA_k} &= \delta_{lk} \\
\label{eqn:wick2} \ev{A_lB_k} &= \frac{2}{\pi} \int_0^\pi d\phi  ( \cos \phi r \left( 1+\lambda\cos\phi \right) - \gamma \lambda \sin\phi \sin\phi r ) \frac{1}{\Lambda_\phi}\\
\label{eqn:wick3}\ev{B_lB_k} &= -\delta_{lk}
\end{align}
in the thermodynamic limit. For finite $L$ equation (\ref{eqn:wick2}) is given by
\begin{equation}
\label{eqn:wick2N}	\ev{A_lB_k} = \frac 1 L \sum_q \frac{1}{\Lambda_q} ( \cos r\phi_q(1+\lambda\cos\phi_q) - \gamma\lambda\sin\phi_q\sin r\phi_q  ).
\end{equation}
To evaluate $\ev{\sigma_i^m\sigma_j^n\sigma_k^o}$ and  $\ev{\sigma_i^m\sigma_j^n\sigma_k^o\sigma_l^p}$ we express the Pauli operators in terms of $A_j$ or $B_k$. Then proceed with the general scheme by Ref.~\cite{Caianiello52}. Reorder the operators with respect to the fermionic commutation relations, such that all $A_l$ are in front of the $B_l$ and both are in ascending order. After this step the expressions look like
\begin{equation}
 \pm\ev{A_{i_1}...A_{i_k}B_{j_1}...B_{j_k}}.
\end{equation}
Applying the Wick theorem iteratively one obtains the following Pfaffian
\begin{equation}
  \pm  pf \begin{bmatrix}
	0 & \ev{A_{i_1}A_{i_2}} & ... & \ev{A_{i_1}A_{i_k}} & \ev{A_{i_1}B_{j_1}}  & ... & \ev{A_{i_1}B_{j_k}}\\
	-\ev{A_{i_1}A_{i_2}} & 0 &  & \ev{A_{i_2}A_{i_k}} & \ev{A_{i_2}B_{j_1}}  & & \ev{A_{i_2}B_{j_k}}\\
	 & &  & \vdots & \vdots & & \vdots\\
	 & & \ddots & \ev{A_{i_{k-1}}A_{i_k}} & \ev{A_{i_{k-1}}B_{j_1}}  & & \ev{A_{i_{k-1}}B_{j_k}}\\
	 & & & 0 & \ev{A_{i_{k}}B_{j_1}} & ... & \ev{A_{i_{k}}B_{j_k}}\\
	 & & & & 0 & ... & \ev{B_{j_1}B_{j_k}}\\
	 & & & & & &  \vdots\\
	 & & & &  & \ddots & \ev{B_{j_{k-1}}B_{j_k}}\\
	 & & & &  & & 0
     \end{bmatrix}.
\end{equation}
Note that the upper left and the lower right $k \times k$ block of this Pfaffian are zero, since we have $i_1<i_2<\dots<i_k$ and $j_1<j_2<\dots<j_k$ together with (\ref{eqn:wick1}) and (\ref{eqn:wick3}). Such a Pfaffian can be expressed as
\begin{equation}
 pf \begin{bmatrix}
0 & M \\
-M^T & 0 
\end{bmatrix} = (-1)^{\frac{k(k-1)}{2}} \det M
\end{equation}
and therefore we have
\begin{equation}
 \ev{A_{i_1}...A_{i_k}B_{j_1}...B_{j_k}} = (-1)^{\frac{k(k-1)}{2}} 
\begin{vmatrix}
	G_{j_1-i_1} & ... & G_{j_k-i_1}\\
	\vdots & & \vdots\\
	G_{j_1 -i_k} & ... & G_{j_k-i_k}
\end{vmatrix}.
\end{equation}
where $ G_{k-l}=\ev{A_lB_k}.$

\subsection{Expectation values of four-site operator basis} 
\label{appendix7}
A list of all non-zero expectation values of the four-site Pauli operators $\ev{\sigma_i^m\sigma_j^n\sigma_k^o\sigma_l^p}$ is presented here. The three site expectation values are omitted, since they are already contained in those of four sites. Note that there are 80 non-vanishing expectation values out of 256. This is due to several symmetries the Hamiltonian possesses, and which carry over to its ground state. Namely the Hamiltonian is translationally invariant and thus just the spacings $\alpha=j-i$, $\beta=k-j$ and $\delta=k-l$ between the local sites enter the final expectation values. Furthermore, it is invariant under global $x$-flip $\sigma_i^x\rightarrow -\sigma_i^x$ for all $i$ and under global $y$-flip. Hence the only contribution to the four site Pauli-operators are those, in which $\sigma^x$ and $\sigma^y$ appear an even number of times. The exact numerical value finally depends on $\ev{A_lB_k} = G_{k-l}$ only.\\
It is
\begin{align}
\nonumber	\ev{\eins \eins \eins \eins} &= 1,\\
\nonumber	\ev{\sigma_i^z \eins \eins \eins} &= -G_0,\\
		\ev{\sigma_i^z \sigma_j^z \eins \eins} &= G_0^2 -G_\alpha G_{-\alpha}.
\end{align}
Similar it is $\ev{\eins \sigma_j^z \eins \eins} = \ev{\eins \eins \sigma_k^z \eins} =\ev{\eins \eins \eins \sigma_l^z} = -G_0 $. As for the tensor product of two $\sigma^z$ operators $ \ev{\sigma_i^z \eins \sigma_k^z \eins} = G_0^2 -G_{\alpha+\beta} G_{-\alpha-\beta}$, $\ev{\sigma_i^z \eins \eins \sigma_l^z } = G_0^2 -G_{\alpha+\beta+\delta} G_{-\alpha-\beta-\delta}$, $\ev{\eins \sigma_j^z \sigma_k^z \eins} = G_0^2 -G_\beta G_{-\beta}$, $ \ev{\eins \sigma_j^z \eins \sigma_j^z} = G_0^2 -G_{\beta+\delta} G_{-\beta-\delta}$, $ \ev{\eins \eins \sigma_k^z \sigma_l^z} = G_0^2 -G_\delta G_{-\delta}$.
\begin{equation}
\ev{\sigma_i^z \sigma_j^z \sigma_k^z \eins} 
  = -\begin{vmatrix}
	G_0 & G_\alpha & G_{\alpha+\beta}\\
	G_{-\alpha} & G_0 & G_\beta\\
	G_{-\alpha-\beta} & G_{-\beta} & G_0
     \end{vmatrix}.
\end{equation}
To get the expectation values of the other possible permutations of the above operator one simply has to perform the following replacements (applied from left to right) $\ev{\sigma_i^z \sigma_j^z \eins \sigma_l^z}: \beta\rightarrow\beta+\delta$,  $\ev{\sigma_i^z \eins \sigma_k^z \sigma_l^z}: \beta\rightarrow\delta \land \alpha\rightarrow\alpha+\beta$ and $\ev{\eins \sigma_j^z \sigma_k^z \sigma_l^z}: \beta\rightarrow\delta \land \alpha\rightarrow\beta$. 
\begin{equation}
\ev{\sigma_i^z \sigma_j^z \sigma_k^z \sigma_l^z}
  = \begin{vmatrix}
	G_0 & G_\alpha & G_{\alpha+\beta} & G_{\alpha+\beta+\delta}\\
	G_{-\alpha} & G_0 & G_{\beta} & G_{\beta+\delta}\\
	G_{-\alpha-\beta} & G_{-\beta} & G_{0} & G_{\delta}\\
	G_{-\alpha-\beta-\delta} & G_{-\beta-\delta} & G_{-\delta} & G_{0}\\
     \end{vmatrix}.
\end{equation}

\begin{equation}
\label{eqn:xx11}
\ev{\sigma_i^x \sigma_j^x \eins \eins} = (-1)^\alpha \begin{vmatrix}
	G_{-1} & ... & G_{\alpha-2}\\
	\vdots & & \vdots\\
	G_{-\alpha} & ... & G_{-1}
\end{vmatrix},
\end{equation}
$\ev{\sigma_i^x \eins \sigma_k^x \eins}: \alpha\rightarrow \alpha+\beta$, $\ev{\sigma_i^x \eins \eins \sigma_l^x }: \alpha\rightarrow \alpha+\beta+\delta$, $\ev{\eins \sigma_j^x \sigma_k^x \eins}: \alpha\rightarrow \beta$, $\ev{\eins \sigma_j^x \eins \sigma_l^x }: \alpha \rightarrow \beta+\delta$, $\ev{\eins \eins \sigma_k^x \sigma_l^x}: \alpha\rightarrow \delta$. 

\begin{equation}
\label{eqn:xxz1}
\ev{\sigma_i^x \sigma_j^x \sigma_k^z \eins} = (-1)^{\alpha+1} \begin{vmatrix}
	G_{-1} & ... & G_{\alpha-2} & G_{\alpha+\beta-1}\\
	\vdots & & \vdots & \vdots\\
	G_{-\alpha} & ... & G_{-1} & G_\beta\\
	G_{-\alpha-\beta} & ... & G_{-\beta-1} & G_0
\end{vmatrix},
\end{equation}
$\ev{\sigma_i^x \sigma_j^x \eins \sigma_l^z}: \beta \rightarrow \beta+\delta$, $\ev{\sigma_i^x \eins \sigma_k^x \sigma_l^z}: \beta \rightarrow \delta \land \alpha \rightarrow \alpha+\beta$, $\ev{\eins \sigma_j^x \sigma_k^x \sigma_l^z}: \beta \rightarrow \delta \land \alpha \rightarrow \beta$.

\begin{equation}
\label{eqn:xzx1}
\ev{\sigma_i^x \sigma_j^z \sigma_k^x \eins} = (-1)^{\alpha+\beta} \begin{vmatrix}
	G_{-1} & ... & G_{\alpha-2} & G_{\alpha} & ... &  G_{\alpha+\beta-2}\\
	\vdots & & \vdots & \vdots & & \vdots \\
	G_{-\alpha+1} & ... & G_{0} & G_{2} & ... &  G_{\beta}\\
	G_{-\alpha-1} & ... & G_{-2} & G_{0} & ... &  G_{\beta-2}\\
	\vdots & & \vdots & \vdots & & \vdots\\
	G_{-\alpha-\beta} & ... & G_{-\beta-1} & G_{-\beta+1} & ... &  G_{-1}\\
\end{vmatrix},
\end{equation}
$\ev{\sigma_i^x \sigma_j^z \eins \sigma_l^x}: \beta \rightarrow \beta+\delta$, $\ev{\sigma_i^x \eins \sigma_k^z \sigma_l^x}: \beta \rightarrow \delta \land \alpha \rightarrow \alpha+\beta$, $\ev{\eins \sigma_j^x \sigma_k^z \sigma_l^x}: \beta \rightarrow \delta \land \alpha \rightarrow \beta$.

\begin{equation}
\label{eqn:zxx1}
\ev{\sigma_i^z \sigma_j^x \sigma_k^x \eins} = (-1)^{\beta+1} \begin{vmatrix}
	G_{0} & G_{\alpha} & ... &  G_{\alpha+\beta-1}\\
	G_{-\alpha-1} & G_{-1} & ... &  G_{\beta-2}\\
	\vdots & \vdots & & \vdots\\
	G_{-\alpha-\beta} & G_{-\beta} & ... &  G_{-1}\\
\end{vmatrix},
\end{equation}
$\ev{\sigma_i^z \sigma_j^x \eins \sigma_l^x}: \beta \rightarrow \beta+\delta$, $\ev{\sigma_i^z \eins \sigma_k^x \sigma_l^x}: \beta \rightarrow \delta \land \alpha \rightarrow \alpha+\beta$, $\ev{\eins \sigma_j^z \sigma_k^x \sigma_l^x}: \beta \rightarrow \delta \land \alpha \rightarrow \beta$.

\begin{equation}
\label{eqn:yy11}
\ev{\sigma_i^y \sigma_j^y \eins \eins} = (-1)^\alpha \begin{vmatrix}
	G_{1} & ... & G_{\alpha}\\
	\vdots & & \vdots\\
	G_{-\alpha+2} & ... & G_{1}
\end{vmatrix},
\end{equation}
$\ev{\sigma_i^y \eins \sigma_k^y \eins}: \alpha\rightarrow \alpha+\beta$, $\ev{\sigma_i^y \eins \eins \sigma_l^y }: \alpha\rightarrow \alpha+\beta+\delta$, $\ev{\eins \sigma_j^y \sigma_k^y \eins}: \alpha\rightarrow \beta$, $\ev{\eins \sigma_j^y \eins \sigma_l^y }: \alpha \rightarrow \beta+\delta$, $\ev{\eins \eins \sigma_k^y \sigma_l^y}: \alpha\rightarrow \delta$. 

\begin{equation}
\label{eqn:yyz1}
\ev{\sigma_i^y \sigma_j^y \sigma_k^z \eins} = (-1)^{\alpha+1} \begin{vmatrix}
	G_{1} & ... & G_{\alpha} & G_{\alpha+\beta}\\
	\vdots & & \vdots & \vdots\\
	G_{-\alpha+2} & ... & G_{1} & G_{\beta+1}\\
	G_{-\alpha-\beta+1} & ... & G_{-\beta} & G_0
\end{vmatrix},
\end{equation}
$\ev{\sigma_i^y \sigma_j^y \eins \sigma_l^z}: \beta \rightarrow \beta+\delta$, $\ev{\sigma_i^y \eins \sigma_k^y \sigma_l^z}: \beta \rightarrow \delta \land \alpha \rightarrow \alpha+\beta$, $\ev{\eins \sigma_j^y \sigma_k^y \sigma_l^z}: \beta \rightarrow \delta \land \alpha \rightarrow \beta$.

\begin{equation}
\label{eqn:yzy1}
\ev{\sigma_i^y \sigma_j^z \sigma_k^y \eins} = (-1)^{\alpha+\beta} \begin{vmatrix}
	G_{1} & ... & G_{\alpha-1} & G_{\alpha+1} & ... &  G_{\alpha+\beta}\\
	\vdots & & \vdots & \vdots & & \vdots \\
	G_{-\alpha+2} & ... & G_{0} & G_{2} & ... &  G_{\beta+1}\\
	G_{-\alpha} & ... & G_{-2} & G_{0} & ... &  G_{\beta-1}\\
	\vdots & & \vdots & \vdots & & \vdots\\
	G_{-\alpha-\beta+2} & ... & G_{-\beta} & G_{-\beta+2} & ... &  G_{1}\\
\end{vmatrix},
\end{equation}
$\ev{\sigma_i^y \sigma_j^z \eins \sigma_l^y}: \beta \rightarrow \beta+\delta$, $\ev{\sigma_i^y \eins \sigma_k^z \sigma_l^y}: \beta \rightarrow \delta \land \alpha \rightarrow \alpha+\beta$, $\ev{\eins \sigma_j^y \sigma_k^z \sigma_l^y}: \beta \rightarrow \delta \land \alpha \rightarrow \beta$.

\begin{equation}
\label{eqn:zyy1}
\ev{\sigma_i^z \sigma_j^y \sigma_k^y \eins} = (-1)^{\beta+1} \begin{vmatrix}
	G_{0} & G_{\alpha+1} & ... &  G_{\alpha+\beta}\\
	G_{-\alpha} & G_{1} & ... &  G_{\beta}\\
	\vdots & \vdots & & \vdots\\
	G_{-\alpha-\beta+1} & G_{-\beta+2} & ... &  G_{1}\\
\end{vmatrix},
\end{equation}
$\ev{\sigma_i^z \sigma_j^y \eins \sigma_l^y}: \beta \rightarrow \beta+\delta$, $\ev{\sigma_i^z \eins \sigma_k^y \sigma_l^y}: \beta \rightarrow \delta \land \alpha \rightarrow \alpha+\beta$, $\ev{\eins \sigma_j^z \sigma_k^y \sigma_l^y}: \beta \rightarrow \delta \land \alpha \rightarrow \beta$.

\begin{equation}
\label{eqn:xxzz}
\ev{\sigma_i^x \sigma_j^x \sigma_k^z \sigma_l^z} = (-1)^{\alpha} \begin{vmatrix}
	G_{-1} & ... & G_{\alpha-2} & G_{\alpha+\beta-1} & G_{\alpha+\beta+\delta-1}\\
	\vdots & & \vdots &\vdots & \vdots\\
	G_{-\alpha} & ... & G_{-1} & G_{\beta} & G_{\beta+\delta}\\
	G_{-\alpha-\beta} & ... & G_{-\beta-1} & G_{0} & G_{\delta}\\
	G_{-\alpha-\beta-\delta} & ... & G_{-\beta-\delta-1} & G_{-\delta} & G_{0}
\end{vmatrix}
\end{equation}

\begin{equation}
\label{eqn:zzxx}
\ev{\sigma_i^z \sigma_j^z \sigma_k^x \sigma_l^x} = (-1)^{\delta} \begin{vmatrix}
	G_0 & G_\alpha & G_{\alpha+\beta} & ... & G_{\alpha+\beta+\delta-1} \\
	G_{-\alpha} & G_0 & G_\beta & ... & G_{\beta+\delta-1}\\
	G_{-\alpha-\beta-1} & G_{-\beta-1} & G_{-1} & ... & G_{\delta-2}\\
	\vdots & \vdots & \vdots & & \vdots\\
	G_{-\alpha-\beta-\delta} & G_{-\beta-\delta} & G_{-\delta} & ... & G_{-1}
\end{vmatrix}
\end{equation}

\begin{equation}
\label{eqn:zxxz}
\ev{\sigma_i^z \sigma_j^x \sigma_k^x\sigma_l^z} = (-1)^{\beta} \begin{vmatrix}
	G_0 & G_\alpha & ... & G_{\alpha+\beta-1} & G_{\alpha+\beta+\delta}\\
	G_{-\alpha-1} & G_{-1} & ... & G_{\beta-2} & G_{\beta+\delta-1} \\
	\vdots & \vdots & &\vdots & \vdots\\
	G_{-\alpha-\beta} & G_{-\beta} & ... & G_{-1} & G_{\delta}\\
	G_{-\alpha-\beta-\delta} & G_{-\beta-\delta} & ... & G_{-\delta-1} & G_{0}
\end{vmatrix}
\end{equation}

\begin{equation}
\label{eqn:xzzx}
\ev{\sigma_i^x \sigma_j^z \sigma_k^z \sigma_l^x} = (-1)^{\alpha+\beta+\delta} \begin{vmatrix}
	G_{-1} & ... & G_{\alpha-2} & G_{\alpha} & ... & G_{\alpha+\beta-2} & G_{\alpha+\beta} & ... & G_{\alpha+\beta+\delta-2}\\
	\vdots & & \vdots & \vdots & & \vdots & \vdots & & \vdots\\
	G_{-\alpha+1} & ... & G_{0} & G_{2} & ... & G_{\beta} & G_{\beta+2} & ... & G_{\beta+\delta}\\
	G_{-\alpha-1} & ... & G_{-2} & G_{0} & ... & G_{\beta-2} & G_{\beta} & ... & G_{\beta+\delta-2}\\
	\vdots & & \vdots & \vdots & & \vdots & \vdots & & \vdots\\
	G_{-\alpha-\beta+1} & ... & G_{-\beta} & G_{-\beta+2} & ... & G_{0} & G_{2} & ... & G_{\delta}\\
	G_{-\alpha-\beta-1} & ... & G_{-\beta-2} & G_{-\beta} & ... & G_{-2} & G_{0} & ... & G_{\delta-2}\\
	\vdots & & \vdots & \vdots & & \vdots & \vdots & & \vdots\\
	G_{-\alpha-\beta-\delta} & ... & G_{-\beta-\delta-1} & G_{-\beta-\delta+1} & ... & G_{-\delta-1} & G_{-\delta+1} & ... & G_{-1}
\end{vmatrix}
\end{equation}

\begin{equation}
\label{eqn:yyzz}
\ev{\sigma_i^y \sigma_j^y \sigma_k^z \sigma_l^z} = (-1)^{\alpha} \begin{vmatrix}
	G_{1} & ... & G_{\alpha} & G_{\alpha+\beta} & G_{\alpha+\beta+\delta}\\
	\vdots & & \vdots &\vdots & \vdots\\
	G_{-\alpha+2} & ... & G_{1} & G_{\beta+1} & G_{\beta+\delta+1}\\
	G_{-\alpha-\beta+1} & ... & G_{-\beta} & G_{0} & G_{\delta}\\
	G_{-\alpha-\beta-\delta+1} & ... & G_{-\beta-\delta} & G_{-\delta} & G_{0}
\end{vmatrix}
\end{equation}

\begin{equation}
\label{eqn:zzyy}
\ev{\sigma_i^z \sigma_j^z \sigma_k^y \sigma_l^y} = (-1)^{\delta} \begin{vmatrix}
	G_0 & G_\alpha & G_{\alpha+\beta+1} & ... & G_{\alpha+\beta+\delta}\\
	G_{-\alpha} & G_0 & G_{\beta+1} & ... & G_{\beta+\delta}\\
	G_{-\alpha-\beta} & G_{-\beta} & G_{1} & ... & G_\delta\\
	\vdots & \vdots & \vdots & & \vdots \\
	G_{-\alpha-\beta-\delta+1} & G_{-\beta-\delta+1} & G_{-\delta+2} & ... & G_1
\end{vmatrix}
\end{equation}

\begin{equation}
\label{eqn:zyyz}
\ev{\sigma_i^z \sigma_j^y \sigma_k^y\sigma_l^z} = (-1)^{\beta} \begin{vmatrix}
	G_0 & G_{\alpha+1} & ... & G_{\alpha+\beta} & G_{\alpha+\beta+\delta}\\
	G_{-\alpha} & G_{1} & ... & G_{\beta} & G_{\beta+\delta} \\
	\vdots & \vdots & &\vdots & \vdots\\
	G_{-\alpha-\beta+1} & G_{-\beta+2} & ... & G_{1} & G_{\delta+1}\\
	G_{-\alpha-\beta-\delta} & G_{-\beta-\delta+1} & ... & G_{-\delta} & G_{0}
\end{vmatrix}
\end{equation}

\begin{equation}
\label{eqn:yzzy}
\ev{\sigma_i^y \sigma_j^z \sigma_k^z \sigma_l^y} = (-1)^{\alpha+\beta+\delta} \begin{vmatrix}
	G_{1} & ... & G_{\alpha-1} & G_{\alpha+1} & ... & G_{\alpha+\beta-1} & G_{\alpha+\beta+1} & ... & G_{\alpha+\beta+\delta}\\
	\vdots & & \vdots & \vdots & & \vdots & \vdots & & \vdots\\
	G_{-\alpha+2} & ... & G_{0} & G_{2} & ... & G_{\beta} & G_{\beta+2} & ... & G_{\beta+\delta+1}\\
	G_{-\alpha} & ... & G_{-2} & G_{0} & ... & G_{\beta-2} & G_{\beta} & ... & G_{\beta+\delta-1}\\
	\vdots & & \vdots & \vdots & & \vdots & \vdots & & \vdots\\
	G_{-\alpha-\beta+2} & ... & G_{-\beta} & G_{-\beta+2} & ... & G_{0} & G_{2} & ... & G_{\delta+1}\\
	G_{-\alpha-\beta} & ... & G_{-\beta-2} & G_{-\beta} & ... & G_{-2} & G_{0} & ... & G_{\delta-1}\\
	\vdots & & \vdots & \vdots & & \vdots & \vdots & & \vdots\\
	G_{-\alpha-\beta-\delta+2} & ... & G_{-\beta-\delta} & G_{-\beta-\delta+2} & ... & G_{-\delta} & G_{-\delta+2} & ... & G_{1}
\end{vmatrix}
\end{equation}

\begin{equation}
\label{eqn:xzxz}
\ev{\sigma_i^x \sigma_j^z \sigma_k^x \sigma_l^z} = (-1)^{\alpha+\beta+1} \begin{vmatrix}
	G_{-1} & ... & G_{\alpha-2} & G_{\alpha} & ... & G_{\alpha+\beta-2} & G_{\alpha+\beta+\delta-1}\\
	\vdots & & \vdots &\vdots & & \vdots & \vdots\\
	G_{-\alpha+1} & ... & G_{0} & G_{2} & ... & G_{\beta} & G_{\beta+\delta+1}\\
	G_{-\alpha-1} & ... & G_{-2} & G_{0} & ... & G_{\beta-2} & G_{\beta+\delta-1}\\
	\vdots & & \vdots & \vdots & & \vdots & \vdots\\
	G_{-\alpha-\beta} & ... & G_{-\beta-1} & G_{-\beta+1} & ... & G_{-1} & G_{\delta}\\
	G_{-\alpha-\beta-\delta} & ... & G_{-\beta-\delta-1} & G_{-\beta-\delta+1} & ... & G_{-\delta-1} & G_0
\end{vmatrix}
\end{equation}

\begin{equation}
\label{eqn:zxzx}
\ev{\sigma_i^z \sigma_j^x \sigma_k^z \sigma_l^x} = (-1)^{\beta+\delta+1} \begin{vmatrix}
	G_0 & G_\alpha & ... & G_{\alpha+\beta-1} & G_{\alpha+\beta+1} & ... & G_{\alpha+\beta+\delta-1}\\
	G_{-\alpha-1} & G_{-1} & ... & G_{\beta-2} & G_{\beta} & ... & G_{\beta+\delta-2}\\
	\vdots & \vdots & & \vdots & \vdots & & \vdots\\
	G_{-\alpha-\beta+1} & G_{-\beta+1} & ... & G_0 & G_2 & ... & G_\delta\\
	G_{-\alpha-\beta-1} & G_{-\beta-1} & ... & G_{-2} & G_0 & ... & G_{\delta-2}\\
	\vdots & \vdots & & \vdots & \vdots & & \vdots\\
	G_{-\alpha-\beta-\delta} & G_{-\beta-\delta} & ... & G_{-\delta-1} & G_{-\delta+1} & ... & G_{-1}
\end{vmatrix}
\end{equation}

\begin{equation}
\label{eqn:yzyz}
\ev{\sigma_i^y \sigma_j^z \sigma_k^y \sigma_l^z} = (-1)^{\alpha+\beta+1} \begin{vmatrix}
	G_{1} & ... & G_{\alpha-1} & G_{\alpha+1} & ... & G_{\alpha+\beta} & G_{\alpha+\beta+\delta}\\
	\vdots & & \vdots &\vdots & & \vdots & \vdots\\
	G_{-\alpha+2} & ... & G_{0} & G_{2} & ... & G_{\beta+1} & G_{\beta+\delta+1}\\
	G_{-\alpha} & ... & G_{-2} & G_{0} & ... & G_{\beta-1} & G_{\beta+\delta-1}\\
	\vdots & & \vdots & \vdots & & \vdots & \vdots\\
	G_{-\alpha-\beta+2} & ... & G_{-\beta} & G_{-\beta+2} & ... & G_{1} & G_{\delta+1}\\
	G_{-\alpha-\beta-\delta+1} & ... & G_{-\beta-\delta-1} & G_{-\beta-\delta+1} & ... & G_{-\delta} & G_0
\end{vmatrix}
\end{equation}

\begin{equation}
\label{eqn:zyzy}
\ev{\sigma_i^z \sigma_j^y \sigma_k^z \sigma_l^y} = (-1)^{\beta+\delta+1} \begin{vmatrix}
	G_0 & G_{\alpha+1} & ... & G_{\alpha+\beta-1} & G_{\alpha+\beta+1} & ... & G_{\alpha+\beta+\delta}\\
	G_{-\alpha} & G_1 & ... & G_{\beta-1} & G_{\beta+1} & ... & G_{\beta+\delta}\\
	\vdots & \vdots & & \vdots & \vdots & & \vdots\\
	G_{-\alpha-\beta+1} & G_{-\beta+2} & ... & G_0 & G_2 & ... & G_{\delta+1}\\
	G_{-\alpha-\beta-1} & G_{-\beta} & ... & G_{-2} & G_0 & ... & G_{\delta-1}\\
	\vdots & \vdots & & \vdots & \vdots & & \vdots\\
	G_{-\alpha-\beta-\delta+1} & G_{-\beta-\delta-2} & ... & G_{-\delta} & G_{-\delta+2} & ... & G_1
\end{vmatrix}
\end{equation}

\begin{equation}
\label{eqn:xxyy}
\ev{\sigma_i^x \sigma_j^x \sigma_k^y \sigma_l^y} = (-1)^{\alpha+\delta} \begin{vmatrix}
	G_{-1} & ... & G_{\alpha-2} & G_{\alpha+\beta} & ... & G_{\alpha+\beta+\delta-1}\\
	\vdots & & \vdots &\vdots & & \vdots\\
	G_{-\alpha} & ... & G_{-1} & G_{\beta+1} & ... & G_{\beta+\delta}\\
	G_{-\alpha-\beta} & ... & G_{-\beta-1} & G_{1} & ... & G_{\delta}\\
	\vdots & & \vdots & \vdots & & \vdots\\
	G_{-\alpha-\beta-\delta+1} & ... & G_{-\beta-\delta} & G_{-\delta+2} & ... & G_{1}
\end{vmatrix}
\end{equation}

\begin{equation}
\label{eqn:yyxx}
\ev{\sigma_i^y \sigma_j^y \sigma_k^x \sigma_l^x} = (-1)^{\alpha+\delta} \begin{vmatrix}
	G_1 & ... & G_\alpha & G_{\alpha+\beta} & ... & G_{\alpha+\beta+\delta-1}\\
	\vdots & & \vdots & \vdots & & \vdots\\
	G_{-\alpha+2} & ... & G_1 & G_{\beta+1} & ... & G_{\beta+\delta}\\
	G_{-\alpha-\beta} & ... & G_{-\beta-1} & G_{-1} & ... & G_{\delta-2}\\
	\vdots & & \vdots & \vdots & & \vdots\\
	G_{-\alpha-\beta-\delta+1} & ... & G_{-\beta-\delta} & G_{-\delta} & ... & G_{-1}
\end{vmatrix}
\end{equation}

\begin{equation}
\label{eqn:xyyx}
\ev{\sigma_i^x \sigma_j^y \sigma_k^y \sigma_l^x} = (-1)^{\alpha+\delta} \begin{vmatrix}
	G_{-1} & ... & G_{\alpha-1} & G_{\alpha+\beta} & ... & G_{\alpha+\beta+\delta-2}\\
	\vdots & & \vdots &\vdots & & \vdots\\
	G_{-\alpha+1} & ... & G_{1} & G_{\beta+2} & ... & G_{\beta+\delta}\\
	G_{-\alpha-\beta} & ... & G_{-\beta} & G_{1} & ... & G_{\delta-1}\\
	\vdots & & \vdots & \vdots & & \vdots\\
	G_{-\alpha-\beta-\delta} & ... & G_{-\beta-\delta} & G_{-\delta+1} & ... & G_{-1}
\end{vmatrix}
\end{equation}

\begin{equation}
\label{eqn:yxxy}
\ev{\sigma_i^y \sigma_j^x \sigma_k^x \sigma_l^y} = (-1)^{\alpha+\delta} \begin{vmatrix}
	G_{1} & ... & G_{\alpha-1} & G_{\alpha+\beta} & ... & G_{\alpha+\beta+\delta}\\
	\vdots & & \vdots &\vdots & & \vdots\\
	G_{-\alpha+1} & ... & G_{-1} & G_{\beta} & ... & G_{\beta+\delta}\\
	G_{-\alpha-\beta} & ... & G_{-\beta-2} & G_{-1} & ... & G_{\delta-1}\\
	\vdots & & \vdots & \vdots & & \vdots\\
	G_{-\alpha-\beta-\delta+2} & ... & G_{-\beta-\delta} & G_{-\delta+1} & ... & G_{1}
\end{vmatrix}
\end{equation}

\begin{equation}
\label{eqn:xxxx}
\ev{\sigma_i^x \sigma_j^x \sigma_k^x \sigma_l^x} = (-1)^{\alpha+\delta} \begin{vmatrix}
	G_{-1} & ... & G_{\alpha-2} & G_{\alpha+\beta-1} & ... & G_{\alpha+\beta+\delta-2}\\
	\vdots & & \vdots &\vdots & & \vdots\\
	G_{-\alpha} & ... & G_{-1} & G_{\beta} & ... & G_{\beta+\delta-1}\\
	G_{-\alpha-\beta-1} & ... & G_{-\beta-2} & G_{-1} & ... & G_{\delta-2}\\
	\vdots & & \vdots & \vdots & & \vdots\\
	G_{-\alpha-\beta-\delta} & ... & G_{-\beta-\delta-1} & G_{-\delta} & ... & G_{-1}
\end{vmatrix}
\end{equation}

\begin{equation}
\label{eqn:yyyy}
\ev{\sigma_i^y \sigma_j^y \sigma_k^y \sigma_l^y} = (-1)^{\alpha+\delta} \begin{vmatrix}
	G_{1} & ... & G_{\alpha} & G_{\alpha+\beta+1} & ... & G_{\alpha+\beta+\delta}\\
	\vdots & & \vdots &\vdots & & \vdots\\
	G_{-\alpha+2} & ... & G_{1} & G_{\beta+2} & ... & G_{\beta+\delta+1}\\
	G_{-\alpha-\beta+1} & ... & G_{-\beta} & G_{1} & ... & G_{\delta}\\
	\vdots & & \vdots & \vdots & & \vdots\\
	G_{-\alpha-\beta-\delta+2} & ... & G_{-\beta-\delta+1} & G_{-\delta+2} & ... & G_{1}
\end{vmatrix}
\end{equation}
\end{widetext}


\begin{thebibliography}{99}

\bibitem{Osterloh02}
 A.~Osterloh, L.~Amico, G.~Falci and R.~Fazio,
 Nature {\bf 416}, 608 (2002).
 
 \bibitem{Osborne02}
 T.~J.~Osborne and M.~A.~Nielsen,
 Phys. Rev. A {\bf 66} 032110 (2002).

\bibitem{VidalLatorre03}
G.~Vidal, J.I.~Latorre, E.~Rico, A.~Kitaev,
Phys. Rev. Lett. {\bf 90}, 227902 (2003). 
 
\bibitem{Wu04}
 L.~A.~Wu, M.~S.~Sarandy, and D.~A.~Lidar,
 Phys. Rev. Lett. {\bf 93}, 250404 (2004).

\bibitem{Roscilde04}
 T.~Roscilde, P.~Verrucchi, A.~Fubini, S.~Haas, and V.~Tognetti,
 Phys. Rev. Lett. {\bf 93}, 167203 (2004).
 

\bibitem{Nataf12}
For recent results see
P. Nataf, M. Dogan, and K. {Le Hur},
 Phys. Rev. A {\bf 86}, 043807 (2012);
 Y. Yao, H.-W. Li, Ch.-M. Zhang, Zh.-Q. Yin, W. Chen, G.- C. Guo, and Zh.-F. Han,
 Phys. Rev. A {\bf 86}, 042102 (2012);
F. Altintas and R. Eryigit,
Ann. Phys. (USA) {\bf 327} 3084 (2012);
L. Lepori, G. {De Chiara}, and A. Sanpera,
 Phys. Rev. B {\bf 87}, 235107 (2013);
V. Azimi Mousolou, C. M. Canali, and E. Sj{\"o}qvist,
 Phys. Rev. A {\bf 88}, 012310 (2013);
for a review see
L. Amico, R. Fazio, A. Osterloh, and V. Vedral,
Rev. Mod. Phys. {\bf 80}, 517 (2008).
 
\bibitem{simulations}
For reviews see 
F. Verstraete, J.I. Cirac, V. Murg,
Adv. Phys. {\bf 57}, 143 (2008);
N. Schuch, arXiv:1306.5551.

\bibitem{Patane07}
D.~Patan\`{e}, R.~Fazio and L.~Amico,
New J. Phys. {\bf 9} 322 (2007).

\bibitem{giampaolo}
S. M. Giampaolo and B.C. Hiesmayr,
Phys. Rev. A {\bf 88}, 052305 (2013).

\bibitem{stasinska}
J. Stasi\'nska, B. Rogers, M. Paternostro, G. De Chiara, and A. Sanpera,
Phys. Rev. A {\bf 89}, 032330 (2014).
 
\bibitem{multipartitespin} 
T.-C. Wei, D. Das, S. Mukhopadyay, S. Vishveshwara, and P.M. Goldbart, 
Phys. Rev. A {\bf 71}, 060305(R) (2005);
O. G\"uhne, G. T\'oth, and H.J. Briegel, 
New J. Phys. {\bf 7}, 229 (2005);
R. Or\'us, 
Phys. Rev. Lett. {\bf 100}, 130502 (2008);
H. S. Dhar, A. Sen(De), and U. Sen,
Phys. Rev. Lett. {\bf 111}, 070501 (2013);
H. S. Dhar, A. Sen(De), and U. Sen,
New J. Phys. {\bf 15}, 013043 (2013).

\bibitem{batle}
J. Batle and M. Casas,
Phys. Rev. A {\bf 82}, 062101 (2010).

 
\bibitem{Srednicki93}
M.~Srednicki,
Phys. Rev. Lett. {\bf 71}, 666 (1993);
C.~Holzhey, F.~Larsen and F.~Wilczek,
Nucl. Phys. B {\bf 424}, 44 (1994);
V.E. Korepin,
Phys. Rev. Lett. {\bf 92}, 096402 (2004);
M.B.~Hastings,
J. Stat. Mech. P08024 (2007);
G. Vitagliano, A. Riera, and J. I. Latorre,
New J. Phys. {\bf 12} 113049 (2010);
P. Calabrese, M. Mintchev, E. Vicari, Eur. Phys. Lett. {\bf 97} 20009 (2012); 
P. Zanardi and L. C. Venuti, J. Stat. Mech. P04023 (2013);
J. Haegeman, T. J. Osborne, H. Verschelde, and F. Verstraete, 
Phys. Rev. Lett. 110, 100402 (2013);
B. Collins, I. Nechita, and K. Zyczkowski,
J. Phys. A: Math. Theor. {\bf 46}, 305302 (2013).

\bibitem{ECPlenio10}
J. Eisert, M. Cramer, M.B. Plenio,
Rev. Mod. Phys. {\bf 82}, 277 (2010).
 
\bibitem{Jungnitsch11}
 B.~Jungnitsch, T.~Moroder and O.~G\"{u}hne,
 Phys. Rev. Lett. {\bf 106}, 190502 (2011).

 \bibitem{Kampermann12}
 J. T. Barreiro, P. Schindler, O. G\"uhne, T. Monz, M. Chwalla, C. F. Roos, M. Hennrich,
 and R. Blatt, Nature Phys. {\bf 6}, 943 (2010);
 H. Kampermann, O. G\"uhne, C. Wilmott, and D. Bru{\ss},
 Phys. Rev. A {\bf 86}, 032307 (2012).
 
 \bibitem{Lieb61}
 E.~Lieb, T.~Schultz and D.~Mattis,
 Ann. Phys. (N.Y.) {\bf 16}, 407 (1961);
 P.~Pfeuty,
 Ann. Phys. (N.Y.) {\bf 57}, 79 (1970);
 E.~Barouch, B.M.~McCoy and M.~Dresden,
 Phys. Rev. A {\bf 2}, 1075 (1970);
 E.~Barouch, B.M.~McCoy and M.~Barry,
 Phys. Rev. A {\bf 3}, 786 (1971).
 
\bibitem{OS05}
A. Osterloh and J. Siewert, Phys. Rev. A {\bf 72}, 012337 (2005).

 \bibitem{gtreview}
 O. G\"uhne and G. T\'oth,
{Phys. Rep.} {\bf 474}, 1 (2009).
 
\bibitem{pptcriterion}
A. Peres, Phys.\ Rev.\ Lett.\ {\bf 77}, 1413 (1996).

\bibitem{pptmixer}
O. G\"uhne, B. Jungnitsch, T. Moroder, and Y.S. Weinstein,
Phys. Rev. A {\bf 84}, 052319 (2011); 
L. Novo, T. Moroder, and O. G\"uhne, 
Phys. Rev. A {\bf 88}, 012305 (2013).
  
 \bibitem{fisher}
 M. E. Fisher, in ``Fenomeni critici'' (ed. by. M.S. Green), Academic Press (New York), 1971.
 
 \bibitem{barber}
 M.N. Barber, in ``Phase transitions and critical phenomena'' (ed. by C. Domb, 
 J.L. Lebowitz), Academic Press (New York), 1983.
 
\bibitem{Uhlmann00}
A. Uhlmann, Phys. Rev. A {\bf 62}, 032307 (2000).

\bibitem{Sen05}
 A.~Sen(De), U.~Sen, and M.~Lewenstein,
Phys. Rev. A {\bf 72}, 052319 (2005).



 \bibitem{pptmixerprogram}
See the program {\tt PPTmixer}, available at
{\tt mathworks.com/matlabcentral/fileexchange/30968}.

\bibitem{Gurvits02}
 L.~Gurvits and H.~Barnum,
 Phys. Rev. A {\bf 66}, 062311 (2002).

\bibitem{Caianiello52}
 E.R.~Caianiello and S.~Fubini,
 Nuovo Cimento {\bf 9} 1218 (1953).

\end{thebibliography}
\end{document}